\documentclass[aps,pra,reprint,amsmath,superscriptaddress,amssymb]{revtex4-1}
\usepackage[ansinew]{inputenc}
\usepackage{graphicx}

\usepackage[colorlinks,linkcolor=black,citecolor=black,urlcolor=black]{hyperref}
\usepackage[all]{hypcap}

\newcommand{\bra}[1]{\langle #1 |}
\newcommand{\ket}[1]{|#1\rangle}

\newcommand{\fref}[1]{Fig.~\ref{fig:#1}}

\usepackage{graphicx}
\usepackage{textcomp}
\usepackage{color}
\usepackage{afterpage}
\usepackage[normalem]{ulem}
\usepackage{pifont}

\begin{document}
\title{Entanglement Distillation between Solid-State Quantum Network Nodes} 
\author{N. Kalb}
\thanks{These authors contributed equally.}
\author{A. A. Reiserer}
\altaffiliation[Present address: ]{Max-Planck-Institute for Quantum Optics, Hans-Kopfermann-Str. 1, 85748 Garching, Germany}
\author{\normalfont\textsuperscript{,*} P. C. Humphreys}
\thanks{These authors contributed equally.}
\author{J. J. W. Bakermans}
\author{S. J. Kamerling}
\affiliation{QuTech, Delft University of Technology, P. O. Box 5046, 2600 GA Delft, The Netherlands}
\affiliation{Kavli Institute of Nanoscience, Delft University of Technology, P. O. Box 5046, 2600 GA Delft, The Netherlands}
\author{N. H. Nickerson}
\affiliation{Department of Physics, Imperial College London, Prince Consort Road, London SW7 2AZ, U.K.}
\author{S. C. Benjamin}
\affiliation{Department of Materials, University of Oxford, Parks Road, Oxford OX1 3PH, U.K.}
\author{D. J. Twitchen}
\author{M. Markham}
\affiliation{Element Six Innovation, Fermi Avenue, Harwell Oxford, Didcot, Oxfordshire OX11 0QE, U.K.}
\author{R. Hanson}
\email[To whom correspondence should be addressed; E-mail: ]{r.hanson@tudelft.nl}
\affiliation{QuTech, Delft University of Technology, P. O. Box 5046, 2600 GA Delft, The Netherlands}
\affiliation{Kavli Institute of Nanoscience, Delft University of Technology, P. O. Box 5046, 2600 GA Delft, The Netherlands}

\begin{abstract}
	The potential impact of future quantum networks hinges on high-quality quantum entanglement shared between network nodes. Unavoidable real-world imperfections necessitate means to improve remote entanglement by local quantum operations. Here we realize entanglement distillation on a quantum network primitive of distant electron-nuclear two-qubit nodes. We demonstrate the heralded generation of two copies of a remote entangled state through single-photon-mediated entangling of the electrons and robust storage in the nuclear spins. After applying local two-qubit gates, single-shot measurements herald the distillation of an entangled state with increased fidelity that is available for further use. In addition, this distillation protocol significantly speeds up entanglement generation compared to previous two-photon-mediated schemes. The key combination of generating, storing and processing entangled states demonstrated here opens the door to exploring and utilizing multi-particle entanglement on an extended quantum network.
\end{abstract}
\maketitle

Future quantum networks connecting nodes of long-lived stationary qubits through photonic channels may enable secure communication, quantum computation and simulation, and enhanced metrology~\cite{kimble_quantum_2008,ben-or_secure_2006,broadbent_universal_2009,jiang_quantum_2009,ekert_ultimate_2014,cirac_distributed_1999,gottesman_longer-baseline_2012,nickerson_freely_2014,komar_quantum_2014}. The power of these applications fundamentally derives from quantum entanglement shared between the network nodes. The key experimental challenge is therefore to establish high-quality remote entanglement in the presence of unavoidable errors such as decoherence, photon loss and imperfect quantum control. Remarkably, by only using classical communication and local quantum operations, a high-fidelity remote entangled state can be distilled from several lower-fidelity copies~\cite{bennett_purification_1996,deutsch_quantum_1996} (\fref{Fig1}A). Success of this intrinsically probabilistic distillation can be non-destructively heralded by measurement outcomes such that the distilled state is available for further use, a critical requirement for scalable networks. Owing to these unique features, entanglement distillation, also known as purification, has become a central building block of quantum network proposals~\cite{cirac_distributed_1999,dur_quantum_1999,gottesman_longer-baseline_2012,komar_quantum_2014,nickerson_freely_2014,childress_fault-tolerant_2006}.

\section*{Generation and distillation of remote entangled states}
To run entanglement distillation on a quantum network, several copies of a raw entangled state must first be shared between the nodes. This can be achieved using a network primitive of two nodes with two qubits each: a communication qubit with an optical interface for generating remote entanglement and a memory qubit for storage (\fref{Fig1}B). First the communication qubits run the entangling protocol, which due to photon loss is intrinsically probabilistic. After photon detection heralds the generation of a raw entangled state on the communication qubits, this state is swapped onto the memory qubits. The communication qubits are then used to generate a second raw entangled state. At this point, the network nodes share two nominally identical copies of the raw state, from which an entangled state of higher fidelity can be distilled. This protocol thus exploits the combination of heralded generation of remote entanglement with robust quantum state storage, high-fidelity quantum logic gates and non-demolition qubit readout within each node.

These demanding experimental requirements have so far limited the exploration of distillation on entangled qubits to four ions within a single node~\cite{reichle_experimental_2006} and to all-photonic protocols without memories in which the distilled state was unavoidably lost upon success~\cite{pan_experimental_2003,kwiat_experimental_2001,pan_multiphoton_2012}. As an important step towards the desired quantum network, heralded entanglement between distant stationary qubits has recently been achieved with ions, atoms, NVs, quantum dots and superconducting qubits~\cite{moehring_entanglement_2007,hofmann_heralded_2012,bernien_heralded_2013,narla_robust_2016,delteil_generation_2016,stockill_phase-tuned_2017}. However, the potential memory qubits investigated so far in conjunction with these protocols~\cite{pfaff_unconditional_2014,hucul_modular_2015} suffered from rapid dephasing during remote entangling attempts due to unwanted couplings, thus precluding the generation of multiple remote entangled states as required for distillation.
\begin{figure*}[hbtp!]
	\centering
	\includegraphics[width=\textwidth]{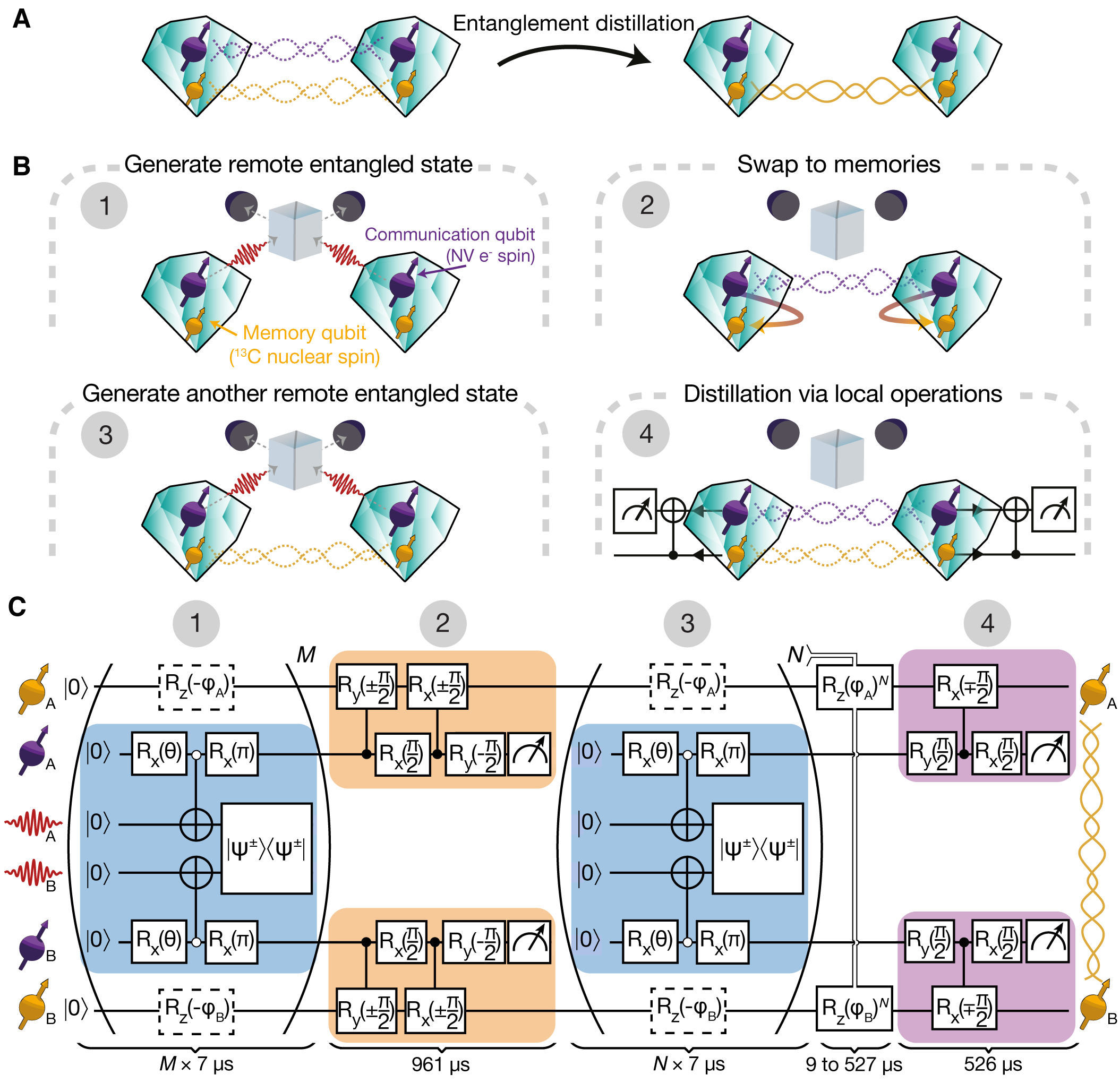}
	\caption{\textbf{Entanglement distillation on a quantum network.} (\textbf{A}) Working principle: a remote entangled state of higher quality (right) is distilled via local operations and classical communication from several lower-quality states (left) that are shared between remote qubits (depicted as colored spins).
		(\textbf{B}) Protocol overview. Each network node consists of a communication qubit (purple) and a memory qubit (yellow). First, the communication qubits are prepared in a remote entangled state by generating entanglement between a photon (red wave packet) and the spin, interfering the optical modes on a beam splitter (gray cube) and subsequently detecting a single photon \ding{172}. Next, the remote entangled state (purple waves represent entanglement) is swapped onto the memory qubits \ding{173}, followed by another round of entangled state generation \ding{174}. Finally, local operations (black circuit) distill a state of higher fidelity \ding{175}. 
		(\textbf{C}) Gate circuit implementing steps \ding{172}-\ding{175}. We include the photonic modes of each setup (red wave packets) in which a qubit is encoded such that vacuum and a single photon represent $\ket{0}$ and $\ket{1}$ respectively. Entanglement between the electron spin and photonic mode is experimentally realized by an optical $\pi$-pulse (depicted as CNOT symbol). The photonic Bell-state projection $\ket{\Psi^\pm}\bra{\Psi^\pm}$ is probabilistically realized by a beam splitter and subsequent detection of a single photon. Dashed-bordered gates indicate phase-shifts of the memory due to free evolution during entangling attempts. Colored boxes indicate logical blocks of the circuit and are used throughout the manuscript.}
	\label{fig:Fig1}
\end{figure*}

We realize the distillation of entangled states on an elementary quantum network consisting of a pair of two-qubit solid-state nodes separated by two meters (\fref{Fig1}B). We achieve this result by implementing a single-photon-based entangling protocol using diamond electron-spin qubits (communication qubits) while capitalizing on recent progress on quantum control~\cite{cramer_repeated_2016} and robust state storage~\cite{reiserer_robust_2016} in nuclear-spin-based quantum memories. Real-time feedback is implemented to compensate memory qubit phase-shifts induced by the probabilistic nature of the remote entangling protocol. As an immediate advantage, the demonstrated protocol distinctly increases the efficiency of entanglement generation compared to the standard two-photon-coincidence protocols used in earlier works~\cite{hucul_modular_2015,hensen_loophole-free_2015}, while removing the optical path-length dependence of stand-alone probabilistic single-photon protocols~\cite{slodicka_atom-atom_2013,delteil_generation_2016}. More generally, by demonstrating the key capabilities for a quantum network in a single experiment, we realize a universal backbone that opens the door to extended quantum networks powered by high-quality remote quantum entanglement.

\section*{Quantum network nodes}
Our implementation of a quantum network node employs a nitrogen-vacancy (NV) electron spin in diamond as a communication qubit and a nearby carbon-13 nuclear spin as a memory qubit. The diamond chips holding these qubits reside in individual closed-cycle cryostats ($T=4\,\mathrm{K}$) that are separated by two meters~\cite{aa_suppmatt_2017}. The electron spin state is manipulated using amplitude-shaped microwave pulses. Electron spin decoherence occurs on timescales exceeding a millisecond and has negligible impact on the presented results.
Spin-selective resonant optical excitation enables high-fidelity initialization and single-shot non-demolition read-out of the electron spin~\cite{blok_manipulating_2014}, as well as generation of spin-photon entanglement for connecting distant nodes~\cite{bernien_heralded_2013}. We employ nuclear spins with intrinsic dephasing times $T_\mathrm{2}^*$ of $3.4(1)$ ms and $16.2(3)$ ms for node A and B, respectively~\cite{aa_suppmatt_2017}. We implement universal control on each of these nuclear spin qubits by exploiting its hyperfine coupling to the electron spin through recently developed dynamical-decoupling-based gate sequences~\cite{taminiau_universal_2014}. This complete quantum toolbox enables the implementation of all four steps in the distillation protocol. 

\fref{Fig1}C shows the compilation of the full gate circuit into the quantum control operations of our platform. This compilation maximizes the repetition rate and minimizes the number of local quantum gates following the generation of the first remote state. In particular, by initializing the memory qubit at the start of the protocol we are able to implement the SWAP operation with just two conditional quantum gates instead of the three that would be required for arbitrary input states~\cite{aa_suppmatt_2017}. Note that our SWAP implementation maps the communication qubit energy eigenstates onto memory superposition states $\ket{\pm X} \equiv (\ket{0}\pm \ket{1})/\sqrt{2}$.

To benchmark the performance of the local quantum logic we execute a combination of the SWAP (yellow box in \fref{Fig1}C) and the gates of the distillation step (purple box in \fref{Fig1}C) to generate a maximally-entangled Bell-state between the communication and memory qubits (see \fref{Fig2}A). The full density matrix of the resulting two-qubit state is reconstructed via quantum state tomography (QST, see ref.~\cite{aa_suppmatt_2017} for further details). We find a fidelity with the ideal Bell state of $0.96(1)$ ($0.98(1))$ for node A (B) indicating high-quality operations in both nodes (\fref{Fig2}B).
\begin{figure}[ht!]
	\centering
	\includegraphics[width=\columnwidth]{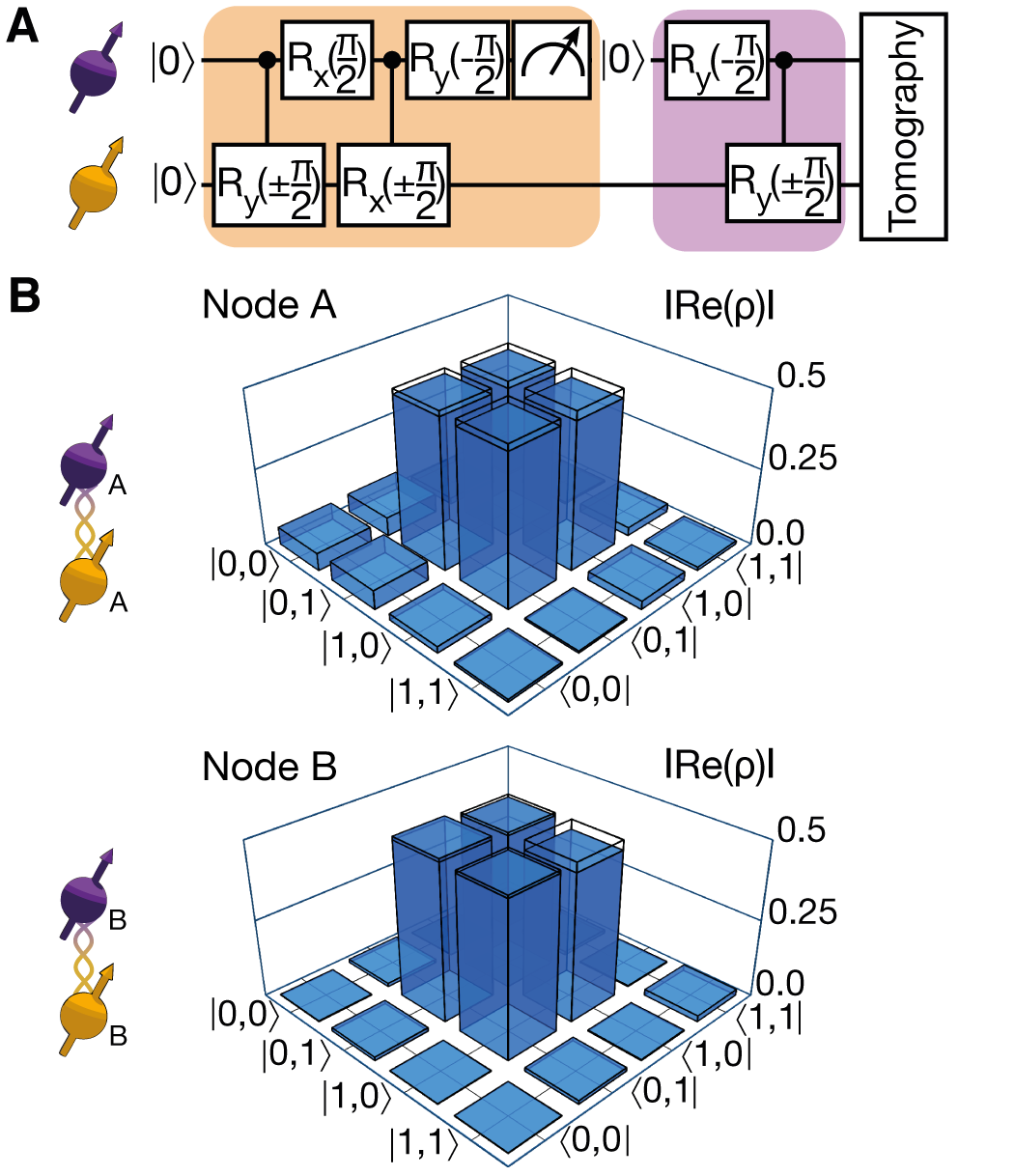}
	\caption{ \textbf{Benchmarking local control.} (\textbf{A}) Gate circuit for entanglement generation within one node. All local operations of the purification circuit are employed to generate a entangled state between communication and memory qubit. Color coding of local operations corresponds to \fref{Fig1}C. (\textbf{B}) Absolute value of the real part of the local density matrix obtained via sequential QST~\cite{aa_suppmatt_2017}. We find fidelities with the desired entangled state of $0.96(1)$ (node A) and $0.98(1)$ (node B). Transparent bars give the values of the ideal state.}
	\label{fig:Fig2}
\end{figure}

\section*{Robust storage of quantum information}

A critical capability for the network nodes is the robust storage of quantum information in the memories while the communication qubits are used to generate remote entangled states. This requires the memory qubits to have long coherence times and be resilient to operations on the communication qubit. The generation of remote entanglement, in particular, poses two challenges as its probabilistic nature means that an \emph{a priori} unknown number of attempts is required.

First, each failed entangling attempt leaves the communication qubits in an unknown state which necessitates a reset by optical pumping. This reset is a stochastic process which, in combination with the always-on hyperfine interaction between communication and memory qubit, causes dephasing of stored memory states~\cite{reiserer_robust_2016,jiang_coherence_2008}. Here we employ memories with a small parallel hyperfine coupling so that the precession frequency of these memories exhibits only a weak dependency $\Delta \omega$ on the state of the communication qubit during the repumping process of a few hundred nanoseconds ($\Delta \omega_\mathrm{A} = 2\pi \cdot 22.4(1)\,\mathrm{kHz}$ and $\Delta \omega_\mathrm{B} = 2\pi \cdot 26.6(1)\,\mathrm{kHz}$). Decoherence via the perpendicular hyperfine component is suppressed by an applied magnetic field of about $40\,\mathrm{mT}$.
\begin{figure}[tb!]
	\centering
	\includegraphics[width=\columnwidth]{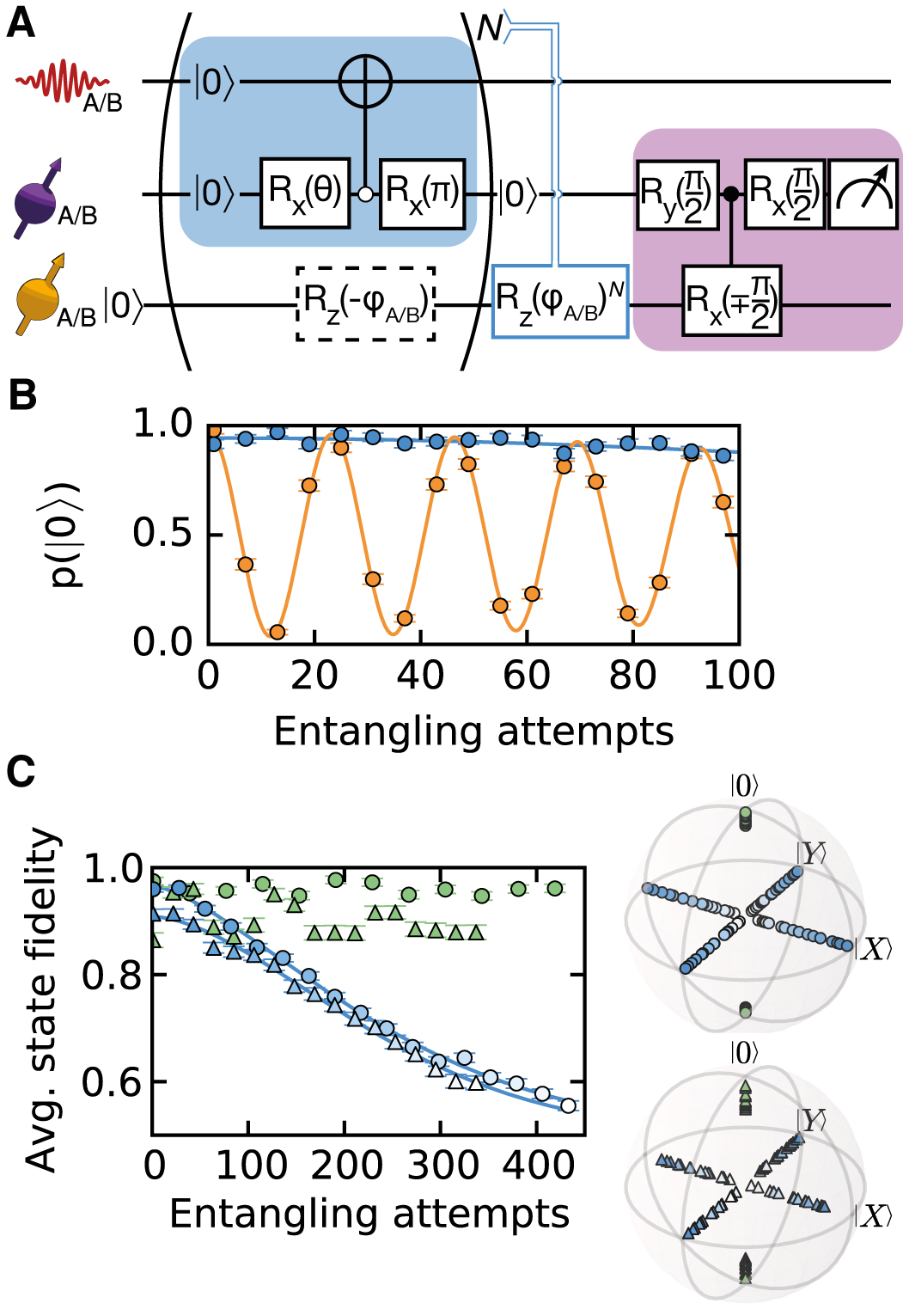}
	\caption{ \textbf{Quantum state storage during entangling operations.} (\textbf{A}) Real-time feedback circuit for memory qubits. We initialize memory A/B, which then experiences a phase-shift of $\varphi_{A/B}$ per executed entangling attempt. After reinitialization of the communication qubit, the distillation step of the protocol is performed (\fref{Fig1}C). Blue-rimmed gate indicates the feedback. (\textbf{B}) Memory state as a function of the number of entangling attempts. The oscillation observed without feedback (orange) is successfully compensated (blue) by the feedback. Solid lines are fits to the data.
		(\textbf{C}) Memory lifetime of node A (triangles) and node B (circles). We initialize the memory in one of the six cardinal states of the Bloch sphere (see right panel, where all non-relevant expectation values are assumed to be zero), sweep the number of entangling attempts, apply feedback, and read-out the relevant expectation value. The average state fidelities are separately plotted for phase-sensitive superposition states (blue) and phase-insensitive eigenstates (green). Blue solid lines depict a generalized exponential fit~\cite{aa_suppmatt_2017}. The decay is limited by the stochastic repumping process, microwave pulse errors and/or environmental dephasing. The color gradient of the left and right panel match to facilitate comparisons. Error bars represent one standard deviation.}
	\label{fig:Fig3}
\end{figure}
Second, the interaction between communication and memory qubit leads to a deterministic phase-shift $\varphi_{A/B}$ on the memory per entangling attempt. Since it is unknown which entangling attempt will herald success, real-time feedback on the memory is required to compensate for these phase-shifts before the final two-qubit gate of \fref{Fig1}C is applied. In addition, the feedback must preserve the coherence of the communication qubit as it holds the second copy of the raw entangled state. We realize such real-time feedback through dynamical decoupling of the electron spin synced with the nuclear spin precession frequency that induces an electron-state-independent phase gate on the memory~\cite{taminiau_universal_2014}. At each node, the number of entangling attempts until success $N$ is tracked by a microprocessor that terminates the subsequent decoupling sequence when the desired rotation $\mathrm{R_z}(\varphi_{A/B})^N$ has been applied. Ideally, this leaves the memory with the desired phase relation regardless of the number of entangling attempts. We calibrate and verify this feedback at each node separately (see \fref{Fig3}A and \fref{Fig3}B) and measure a negligible effect on the memory state fidelity while the state of the communication qubit is preserved as desired.

With this feedback realized, we investigate the robustness of the memory as a function of the elapsed entangling attempts. We initialize the memory in one of the six cardinal states of the Bloch sphere ($\ket{0}$,$\ket{1}$,$\ket{\pm \mathrm{X}}$ and $\ket{\pm \mathrm{Y}} \equiv (\ket{0} \pm i\ket{1})/\sqrt{2}$), execute a number of entangling attempts followed by phase-feedback and measure the relevant memory expectation value (see \fref{Fig3}C). We observe that dephasing-sensitive states $\ket{\pm \mathrm{X}},\ket{\pm \mathrm{Y}}$ decay with $1/e$-values of $273(5)$ ($272(4)$) entangling attempts in node A (node B) whereas the energy eigenstates $\ket{0},\ket{1}$ are preserved with high fidelity as expected. The memories thus provide faithful storage during remote entangling attempts.

\section*{Experimental entanglement distillation}
With local control and storage in place, we now turn towards the execution of the full distillation protocol. Following Ref. \cite{campbell_measurement-based_2008}, we generate the remote states that provide the resources for distillation by first initializing both communication qubits in a superposition with variable angle $\theta$, $\ket{\theta} \equiv \sin\theta\ket{0}-i\cos\theta\ket{1}$. Subsequent optical excitation for state $\ket{0}$ and overlap of the emission of both communication qubits on a beam splitter (see steps $1$ and $3$ of \fref{Fig1}C) generates the raw remote state $\rho_\mathrm{raw}$ if a single photon is detected \cite{campbell_measurement-based_2008}. For equal and small detection probabilities for both nodes and negligible dark counts, $\rho_\mathrm{raw}$ reads:

\begin{equation}
\rho_\mathrm{raw} =  \left(1-\sin^2 \theta \right)\ket{\Psi^\pm_\phi}\bra{\Psi^\pm_\phi}+
\sin^2 \theta \,\ket{0,0}\bra{0,0}.
\label{eq:rawstate}
\end{equation}

The states $\ket{\Psi^\pm_\phi}\equiv (|01\rangle \pm e^{i\phi} |10\rangle ) / \sqrt{2} $ are entangled states, with a relative phase depending on which detector clicked ($\pm$) and an additional internal phase $\phi$ due to the unknown path length between both emitters and the beam splitter. The fraction of the non-entangled admixture $\ket{0,0}\bra{0,0}$ can be directly controlled through the choice of the initial communication qubit state $\ket{\theta}$; note that the choice of $\ket{\theta}$ also affects the probability of successful entanglement generation (scaling as $\sin^2\theta$). We next swap the raw state onto the memories such that the communication qubit is free for another round of remote state generation (step $2$ in \fref{Fig1}C). Once a second state is successfully generated (step $3$), we apply a conditional quantum gate within each node and read out the communication qubits in a single shot. Owing to the quantum non-demolition nature of this readout the memory qubits do not experience additional dephasing during this step \cite{blok_manipulating_2014}. Readout of the communication qubit projects the memories into one of four states depending on the readout results~\cite{aa_suppmatt_2017}:

\[\def\arraystretch{1.5}
\begin{array}{rl}
(0_A,0_B):& \frac{1}{2} \,\cos^4\theta \,U\ket{\Psi^\pm_0}\bra{\Psi^\pm_0}U^\dagger,   \\
(0_A,1_B):&  \frac{1}{2} \,\sin^2\theta \cos^2\theta \, U(\ket{0,1}\bra{0,1}+\ket{1,1}\bra{1,1})U^\dagger,\\ 
(1_A,0_B):&  \frac{1}{2} \,\sin^2\theta \cos^2\theta \,U (\ket{1,0}\bra{1,0}+\ket{1,1}\bra{1,1})U^\dagger,\\
(1_A,1_B):&U\left(\sin^4\theta\ket{1,1}\bra{1,1} +  \frac{1}{2} \cos^4\theta  \ket{\Psi^\pm_{2\phi}}\bra{\Psi^\pm_{2\phi}}\right)U^\dagger.
	\label{eq:PurifyOutcome}
\end{array}
\]

Here the states are left unnormalized; their traces indicate their probabilities of occurrence. The unitary $U$ corresponds to a Hadamard gate on each memory that arises from the swapping operation in each node. Observation of the readout combination ($0_A$,$0_B$) heralds successful distillation and leaves the system in the state
\begin{equation}
\ket{\psi}_\mathrm{c}\otimes \ket{\psi}_\mathrm{m} = e^{i\phi}\ket{0,0} \otimes U\ket{\Psi^\pm_0}
\end{equation} 
with the relative phase of the final Bell state given by the photon detection signature, i.e. the photons in step 1 and 3 were detected in the same (+) or in different (-) output ports. Importantly, the protocol is agnostic to correlated dephasing of the raw states and is therefore only sensitive to optical path length drifts that occur within an individual run of the protocol~\cite{campbell_measurement-based_2008}. This is in stark contrast to probabilistic single-photon protocols~\cite{cabrillo_creation_1999,slodicka_atom-atom_2013,delteil_generation_2016,stockill_phase-tuned_2017} that require path length stabilization over the full course of data acquisition.

The experimental implementation of entanglement generation requires that the communication qubits' optical transitions are kept on resonance despite shot-to-shot fluctuations and long-term drifts of the respective local charge environments. We employ an automatic feedback loop and resonance search routine to compensate for charge jumps such that the experiment is push-button and runs without human intervention. To further optimize the data rate we bound the number of remote entangling attempts to 1000 for step 1 and up to 500 rounds for step 3, leading to event rates (i.e. two remote states were successfully generated) of around 10~Hz. These bounds are a compromise between maximizing the success probability (favoring more attempts) and minimizing effects of drifts and of memory decoherence (favoring fewer attempts).

\section*{Distillation results}
We start by running the complete protocol using $\theta=\pi/6$ and perform full quantum state tomography on the distilled state. This way, using the complete information obtained on the resulting output, we can verify whether the protocol works as desired. \fref{Fig4}A shows the resulting data for a maximum of $50$ entangling attempts in the second round of state generation. This truncation yields optimal state storage during each run of the protocol~\cite{aa_suppmatt_2017}. Quantitatively, the measured fidelity with the ideal Bell state of $0.65(3) > 0.5$ proves entanglement of the distilled state. Furthermore, the density matrix has high populations in the Bell-state-subspace only, showing that the distillation successfully diminishes the separable admixture.

To gain further insight into the performance of the protocol we measure the fidelity of the distilled state for different amounts of the separable admixture in the raw states; i.e. for different $\theta$ (see \fref{Fig4}B, blue dots). The results are again truncated after a maximum of $50$ entangling attempts in the second round. The state fidelities are averaged over both detection signatures. 

\begin{figure} [htbp!]
	\centering
	\includegraphics[width=\columnwidth]{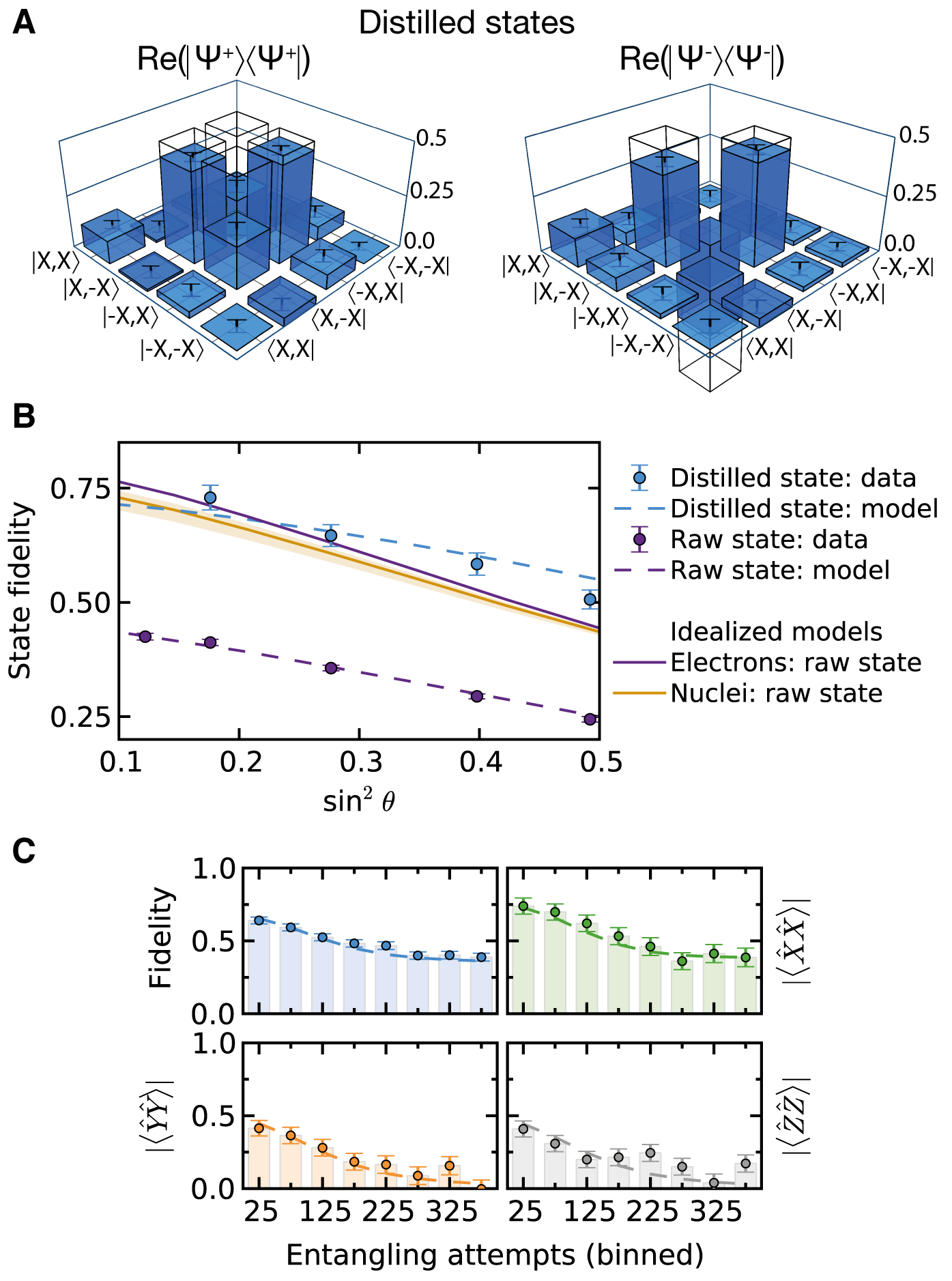}
	\caption{\textbf{Experimental realization of entanglement purification.} (\textbf{A}) Two-qubit density matrices for $\theta = \pi/6$ and a maximum of 50 entangling attempts in the second round. Right panel: different detectors clicked. Left panel: the same detector clicked twice. We find a fidelity with the ideal state of $0.65(3)$ for both states. Transparent bars represent the ideal state. (\textbf{B}) Fidelity with the ideal state as a function of $\theta$ for a maximum of $50$ entangling attempts in the second round. Blue data is the two-memory state fidelity. Dashed lines are derived from our model~\cite{aa_suppmatt_2017}. Purple data is the measured raw state fidelity on the communication qubits. Solid orange (purple) line is the modeled fidelity of the raw state on the memories (communication qubits) that would be obtained if the initial internal phase was known. The memory state is calculated for the average number of entangling attempts until success ($25$). The orange shaded region is the modeled memory fidelity for minimal ($0$ attempts) and maximal ($50$ attempts) dephasing. Fidelities were obtained by measuring the expectation values $\langle \hat{X}\hat{X} \rangle$, $\langle \hat{Y}\hat{Y} \rangle$ and $\langle \hat{Z}\hat{Z} \rangle$. We denote the Pauli operators as $ \hat{X}$, $\hat{Y}$ and $\hat{Z}$. (\textbf{C})  State decay for $\theta = \pi/6$. Data are binned according to the number of second entanglement generation attempts until success. Shown are the state fidelity (blue) and the absolute value of the relevant expectation values. The dashed lines are derived from our theoretical model~\cite{aa_suppmatt_2017}. Error bars represent one standard deviation.}
	\label{fig:Fig4}
\end{figure}

The hallmark of successful distillation is an increase in fidelity of the distilled state compared to that of the raw states. Whereas in the textbook description both raw states are assumed to be equal, in our experiment they are different due to imperfections in the swap operation and memory storage that only affect the raw state held by the memories, and path length variations on short timescales that only affect the raw state held by the communication qubits. To make a meaningful comparison we therefore consider the state fidelities of each of these raw states separately.  

Because of the unreferenced internal phase of the raw states, all coherences are washed out due to optical path length variations. Direct tomography will therefore yield state fidelities that cannot surpass $0.5$. As a result, the measured fidelities of the distilled states far exceed the electron state fidelities measured after step 1 (\fref{Fig4}B, orange dots). Although these numbers reflect the current experiment, we now turn to a more strict comparison by taking into account that the internal phase may become accessible in future experiments through optical path stabilization.

We model the raw state fidelities at the start of the distillation step (step 4) using independently determined parameters under the assumption of a perfectly known initial path length difference (\fref{Fig4}B, solid purple line for raw state on the electrons and solid orange line for raw state on the nuclei) \cite{aa_suppmatt_2017}. For small values of $\theta$ (small separable admixtures) the fidelity increase due to distillation is offset by the errors introduced with the additional quantum operations of the distillation step. However, we find that for larger values of $\theta$ the distilled state fidelity significantly surpasses both of the raw state fidelities (see~\cite{aa_suppmatt_2017} for hypothesis test). This result demonstrates the realization of entanglement distillation on our elementary quantum network. 

For a more detailed understanding of the different error sources contributing to the measured fidelity, we develop an extensive model of the full protocol using independently measured quantities and two free parameters: one factor accounting for additional memory control errors and the second for phase fluctuations of the raw states~\cite{aa_suppmatt_2017}. We find good agreement between the modeled fidelity and the data for each of the different separable admixtures (see \fref{Fig4}B blue dashed line and~\cite{aa_suppmatt_2017}) and for the evolution of the correlations with number of entangling attempts (\fref{Fig4}C). The model indicates that the state fidelities are mainly limited by memory qubit dephasing and control errors as well as non-zero two-photon distinguishability. The latter effect, quantified by a measured two-photon interference visibility of 0.73(3)~\cite{aa_suppmatt_2017}, is especially harmful in the above comparison with the raw state, as this occurs twice for the distillation protocol but only once for the raw state generation. A visibility of 0.9 as observed on different NV center pairs~\cite{hensen_loophole-free_2015} would thus yield an even stronger entanglement enhancement.

\section*{Ebit rate}
Previously demonstrated entangling protocols based on two-photon coincidences~\cite{hucul_modular_2015,hensen_loophole-free_2015} require steps 1 and 3 to succeed in subsequent attempts leading to a success probability scaling with the square of the photon detection probability $p_{\mathrm{det}}$. In contrast, the distillation protocol allows step 3 to succeed in one of many attempts following success in step 1, leading to a success probability scaling linearly with $p_{\mathrm{det}}$ in the ideal case. Given that in a typical quantum network setting $p_{\mathrm{det}}$ will be small (in our case $p_{\mathrm{det}} \approx 10^{-3}$), the distillation protocol can provide a distinct rate advantage despite the overhead of the additional local quantum logic.

\begin{figure}[tbp!]
	\centering
	\includegraphics[width=\columnwidth]{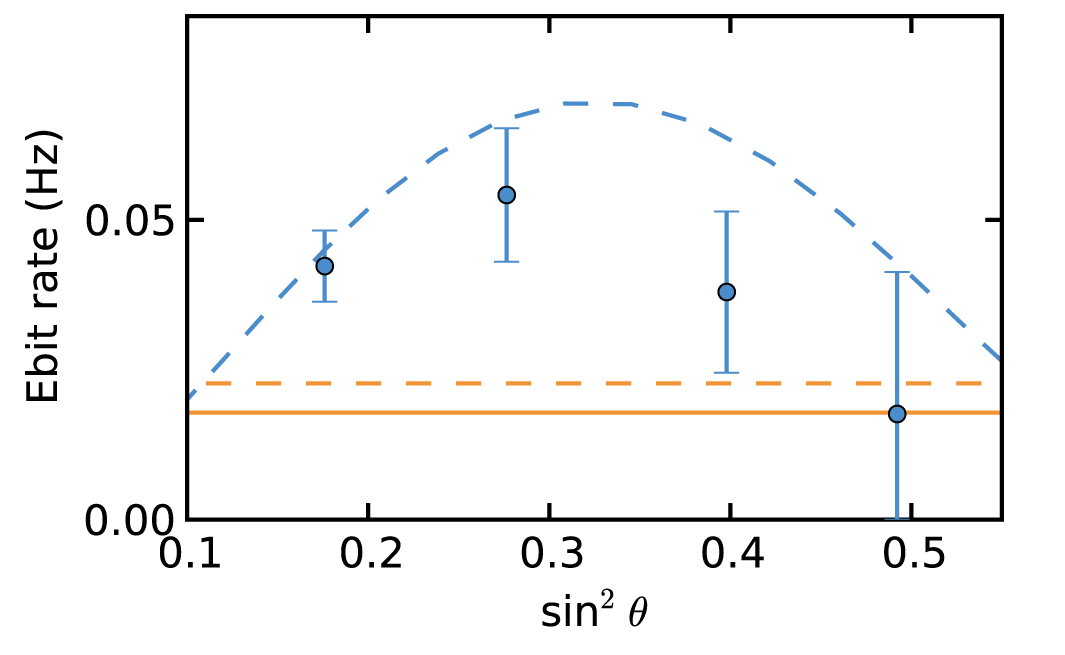}
	\caption{\textbf{Ebit rate comparison.} Ebit rate as a function of excitation angle $\theta$. The blue data is derived from the measured success rates and state fidelity over the entire data set~\cite{aa_suppmatt_2017}. The blue dashed line is the estimated ebit rate for distillation including the overhead of local operations. The solid (dashed) orange line gives the estimated ebit rate of a standard two-photon protocol including (excluding) imperfections. All data are averaged over both detection signatures. Error bars represent one standard deviation.}
	\label{fig:Fig5}
\end{figure}

To quantitatively compare our results with two-photon-coincidence protocols we upper bound the rate $r$ of entangled bit (ebit) generation for each protocol using $r = \nu E_N$ with the logarithmic negativity $E_N$ and the rate of success $\nu$. \fref{Fig5} compares the ebit rate of the presented distillation protocol to the modeled rate of the Barrett-Kok two-photon-coincidence protocol~\cite{barrett_efficient_2005} used in earlier experiments on NV centers~\cite{bernien_heralded_2013,hensen_loophole-free_2015}. We find that the distillation protocol (blue dots) outperforms the two-photon-coincidence protocol for identical experimental conditions, not only when assuming the measured two-photon indistinguishability (orange solid line) but even for the case that the two-photon-coincidence protocol would be able to access perfect two-photon indistinguishability (orange dashed line).

\section*{Conclusion and outlook}
The combination of generating, storing and processing remote entangled qubits as demonstrated in the current distillation experiment provides a universal primitive for realizing extended quantum networks. The distillation itself is a powerful method to counteract unavoidable decoherence as entanglement is distributed throughout the network. Also, the protocol enables a speedup of entanglement generation that can be harnessed in related platforms such as other solid-state defect centers~\cite{christle_isolated_2015} and trapped ions~\cite{monroe_scaling_2013}. Future improvements can be achieved by encoding qubits into decoherence-protected subspaces~\cite{reiserer_robust_2016}, by using isotopically purified materials with longer qubit dephasing times~\cite{balasubramanian_ultralong_2009,maurer_room-temperature_2012,tyryshkin_electron_2012}, by implementing a faster reset or a measurement-based reset of the communication qubit and by increasing the entangling rates through photonic cavities~\cite{faraon_resonant_2011,sipahigil_integrated_2016}. Furthermore, the techniques employed in recent demonstrations of multi-qubit control and quantum error correction on a 4-qubit node~\cite{cramer_repeated_2016,kalb_experimental_2016} are fully compatible with the current experiment, thus highlighting the potential for scaling to more qubits and extending network functionality in the near future. Finally, the methods developed here open the door to exploration and utilization of many-particle entanglement on a multi-node quantum network.

\bibliography{sciencebib}
\bibliographystyle{Science}
\begin{acknowledgements}
	$\quad$ \newline
	We thank Earl Campbell, David DiVincenzo, Mikhail Lukin, Tracy Northup, Lieven Vandersypen and Stephanie Wehner for fruitful discussions. We acknowledge support from the EPSRC National Quantum Technology Hub in Networked Quantum Information Technology (SCB), the Netherlands Organisation for Scientific Research (NWO) through a VICI grant (RH), and the European Research Council through a Starting Grant (RH).
\end{acknowledgements}
\end{document}